\documentstyle[11pt,paspconf,epsf]{article} 
\begin{document}
\title{Creating Spectral Templates from Multicolor Redshift Surveys} 

\author{T.  Budav\'ari\altaffilmark{1}, A.S.  Szalay,
	A.J. Connolly\altaffilmark{2}, I.  Csabai\altaffilmark{1},
	M.E. Dickinson\altaffilmark{3}}

\affil{Department of Physics and   Astronomy, The Johns  Hopkins  University,
Baltimore, MD 21218}

\author{and the HDF/NICMOS Team} 

\altaffiltext{1}{Department of Physics, 
		E\"otv\"os Lor\'and University, Budapest, H-1088}
\altaffiltext{2}{Department of Physics and Astronomy, 
		University of Pittsburgh, Pittsburgh, PA 15260}
\altaffiltext{3}{Space Telescope Science Institute, Baltimore, MD 21218}

\begin{abstract}
We present a  novel method  capable  of  creating optimal  eigenspectra  from
multicolor  redshift   surveys   for  photometric redshift   estimation.  Our
iterative  training  algorithm modifies  the     templates to represent   the
photometric measurements better.    We present a   short description  of  our
algorithm here.   We  show that the   corrected  templates give more  precise
photometric redshifts,  essentially  a ``free'' feature,  since we  were  not
fitting for the redshifts themselves.
\end{abstract}


\section{Introduction}

In the last couple of years photometric redshift estimation has become a very
powerful tool in getting statistical information about  our universe and also
frequently  used  for spectroscopic target   selection to  measure spectra of
faint  galaxies.  Different techniques  have been developed  since Baum (`62)
but all of them may be classified into two groups  based on the approach they
use.

The first are the  so called {\it empirical} methods  (Connolly et al. `95a).
Having a set   of objects with  spectroscopic and  photometric  data, one can
easily  obtain  an  analytic, usually polynomial    fitting function for  the
redshift  over the color ``hyperplane''.   Lower order  piecewise fits can be
also applied  efficiently  (Brunner  et al. `99).  The  function  can be very
quickly evaluated  at any photometric  data point.  However,  this method has
some disadvantages.  We  need to have  a very  good training  set to  get the
fitting formula  and we still will  not be able  to extrapolate  far from the
training set.  The fitting formula only works for a specific set of passbands
and limited redshift ranges, thus if we want to get photometric redshifts for
different catalogs, we need training  sets for all the photometric standards,
so it is very hard to get consistent redshift estimations for them.

In the  {\it template} or  SED  (spectral energy distribution) {\it  fitting}
methods  a set of restframe  spectra are used to work  out  what spectrum and
redshift give  better representation of the color   of a galaxy.   Using this
approach gives us not just the redshift of the galaxy but also an approximate
spectral  type  information.    The  technique works very  well   on separate
photometric catalogs with different  filter sets but it  is obvious  that the
basis spectra  are crucial.  The  currently  used template spectra  come from
both synthetic (Bruzual \& Charlot `93) and measured  (Coleman, Wu \& Weedman
`80) spectral libraries but generally they are  applied as they come with the
hope that they are good enough.  An advanced  version of the template fitting
algorithm has  been developed to  use  ``continuum number'' of  spectra using
eigenspectra instead of a large number of  actual spectra in a very efficient
way in terms of  CPU time and memory.  Using   eigenspectra (Connolly et  al.
`95b) means we consider all spectra to  be approximately a linear combination
of a small number of spectra,  the eigenspectra.  It also  turns out there is
an   optimal  subspace  filtering method  (see   AJC's  talk; Budav\'{a}ri et
al. `99) which gives more reliable restframe  spectra than direct coefficient
fitting. 

In this paper we   describe a new method   which has been developed  to bring
together    the advantages   of both  techniques.    We  show   how to  train
eigenspectra in an iterative procedure to represent the photometry better. In
the  training procedure different catalogs  and also measured  spectra can be
incorporated.   The resulting  eigentemplates     can be used  for   redshift
estimation on any photometry catalog and the redshift prediction becomes more
accurate.

\section{Template Reconstruction}

Our goal is to find  out the underlying basis   template spectra that can  be
used later for photometric redshift estimation.  Since we observe galaxies at
different redshifts,   their  restframe  spectra  are  sampled  at  different
wavelengths by the blueshifted filters.  If  we have a deep enough multicolor
redshift  survey  (or more) then the   rough SEDs defined  by the blueshifted
passbands  are overlapping with each other.   This means the eigenspectra are
oversampled and this  is the fact which allows  us to derive applications for
extracting high resolution  eigentemplates from  broadband photometry (Csabai
et al. `99).  Our new approach  is  different mainly  in spectrum repairation
that we introduce in the next section.

The  iterative training procedure follows a  fully statistical approach.  Its
robustness is provided by the Karhunen-Lo\`eve  (KL) expansion (Karhunen `47;
Lo\`eve   `48).  This transformation  has  been used to derive eigentemplates
from spectra and this is just what we do in our iterative training.  Starting
from a  set of eigenspectra we  can compute a best fitting  type for each and
every galaxy in  our photometric catalog based  on the  known redshift.  This
gives us an approximation to the  restframe spectra of  the galaxies.  Now we
can check  how good  the  spectrum  is  at  least  at such wavelengths  where
photometric data are  available.  We modify ---  repair --- ``smoothly''  the
spectra to represent the measured values better.  This spectrum correction is
the only tricky step that we will describe later on.  Having all the modified
spectra  we   can invoke the  KL   transformation  and  obtain  a  new set of
eigentemplates.  The iteration can be continued until the correlation between
simulated and measured flux values is strong enough.

\paragraph{Repairing Spectra.} 

The way we modify the  spectrum of a  galaxy is the key  step of our training
algorithm.  The best fitting linear  combination  of the eigenspectra is  the
current approximation of the galaxy  that we are  about to correct due to the
photometric values.  This is a  multidimensional minimization problem for the
spectrum itself, where the   dimension of the  problem  is determined  by the
spectral resolution.  The cost function is built up  by two terms.  The first
part corresponds to the deviation  of the spectrum  based on simulated fluxes
from the  measured photometric data  and the second part  is the deviation of
the spectrum from the template based linear combination.
\begin{equation}
\chi^2(\vec s)  = \sum_n {1 \over \Delta_n^2}  \{f_n - \hat f_n(\vec s)\}^2 +
	       \sum_k {1 \over \sigma_k^2} \{s_k - s^t_k\}^2 
\end{equation}
$\vec s$ is the discrete  representation  of the  spectrum  and $s_k$ is  the
$k$th element  of $\vec s$, where $k$  refers to the wavelength, $\lambda_k$.
$s^t_k$  stands for the template based  spectrum and $\sigma_k$ describes the
ability of the spectrum to be changed at a given wavelength.  The photometric
error in the $n$th passband is  represented by $\Delta_n$.  This minimization
problem can be reduced to a system of linear  equations, since $\hat f_n(\vec
s)$ is linear in its variables, $\{s_k\}$; so  it is easy  to solve.  The new
KL basis built from the modified spectra will span a different subspace which
represents    the photometry better.   The details    of this  method will be
described elsewhere (Budav\'{a}ri et al. `99).

\begin{figure}
\plottwo{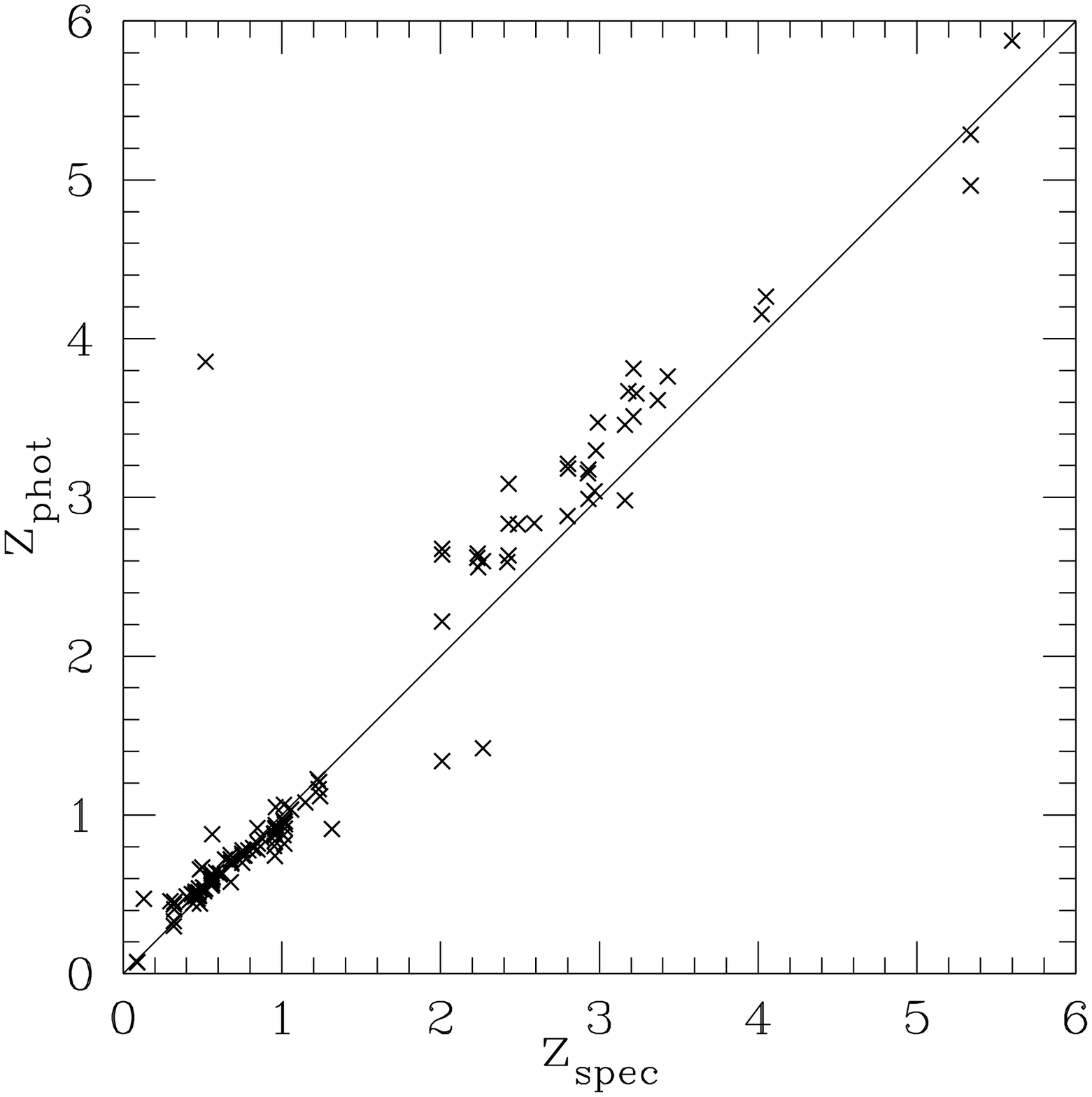}{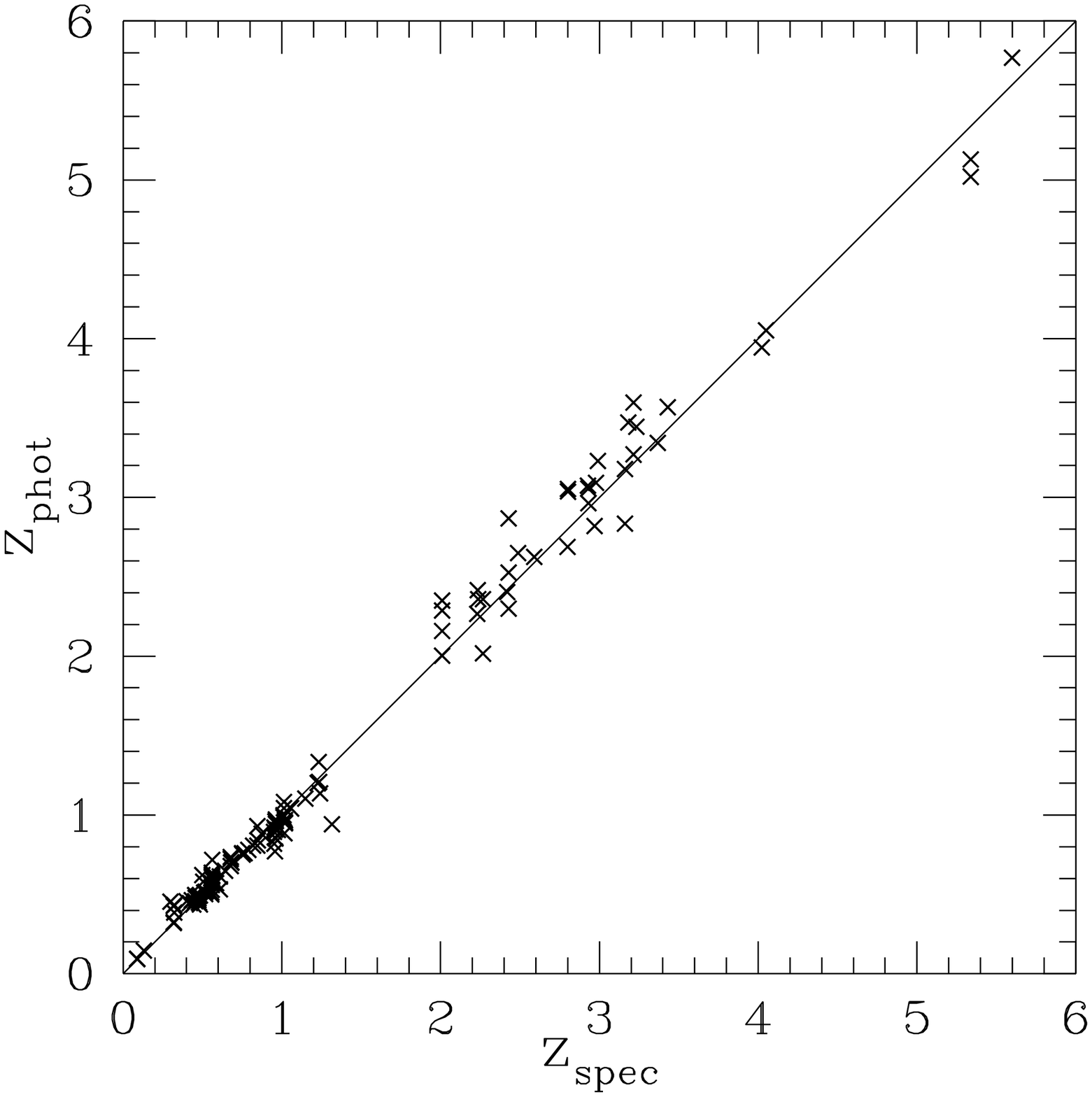}
\caption{Correlation between spectroscopic and photometric redshifts.  On the
right - using 3 CWW eigenspectra, left - using 3 KL eigenspectra after 30
iterations. See Figure \ref{eifig}. for corresponding
eigenspectra. \label{zzfig}}
\end{figure}

\section{Application}

Our training  algorithm has been applied to  the HDF/NICMOS catalog (Williams
et al.  `96; Dickinson et al. `99) which  has unique photometric data quality
and there  is a reasonable   number  of objects  with spectroscopic  redshift
(Cohen, `98).  There are photometric data available in seven passbands: 4 HDF
filters (F300W,  F450W, F606W, F814W), 2  NICMOS filters (F110W, F160W) and a
ground-based K'.   Our initial eigentemplates were  derived from the Coleman,
Wu  \& Weedman (CWW) spectra that  usually provide better redshift estimation
than others (Fernandez-Soto et al., `99; Hogg et al., `98).

\begin{figure}
\plottwo{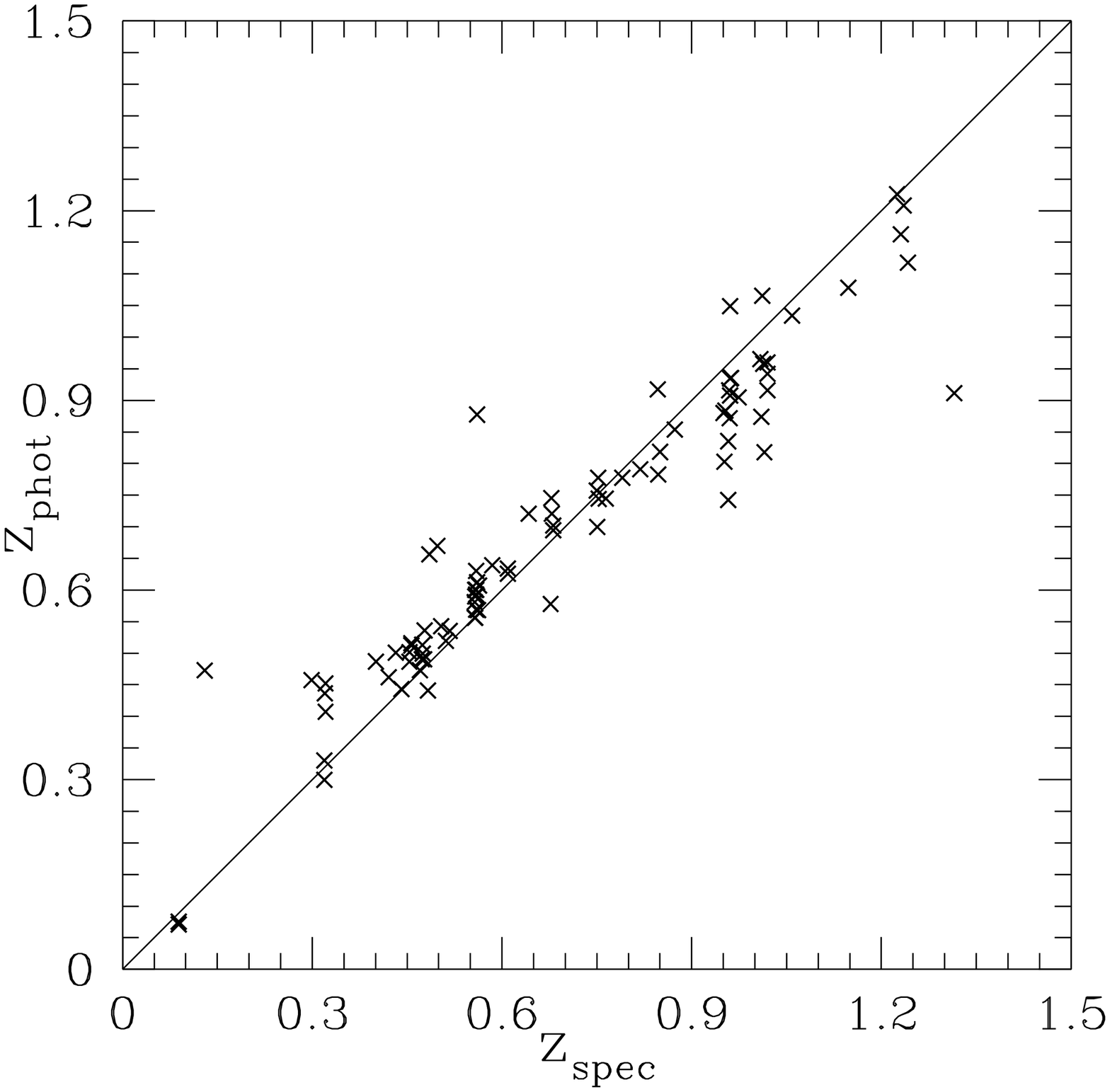}{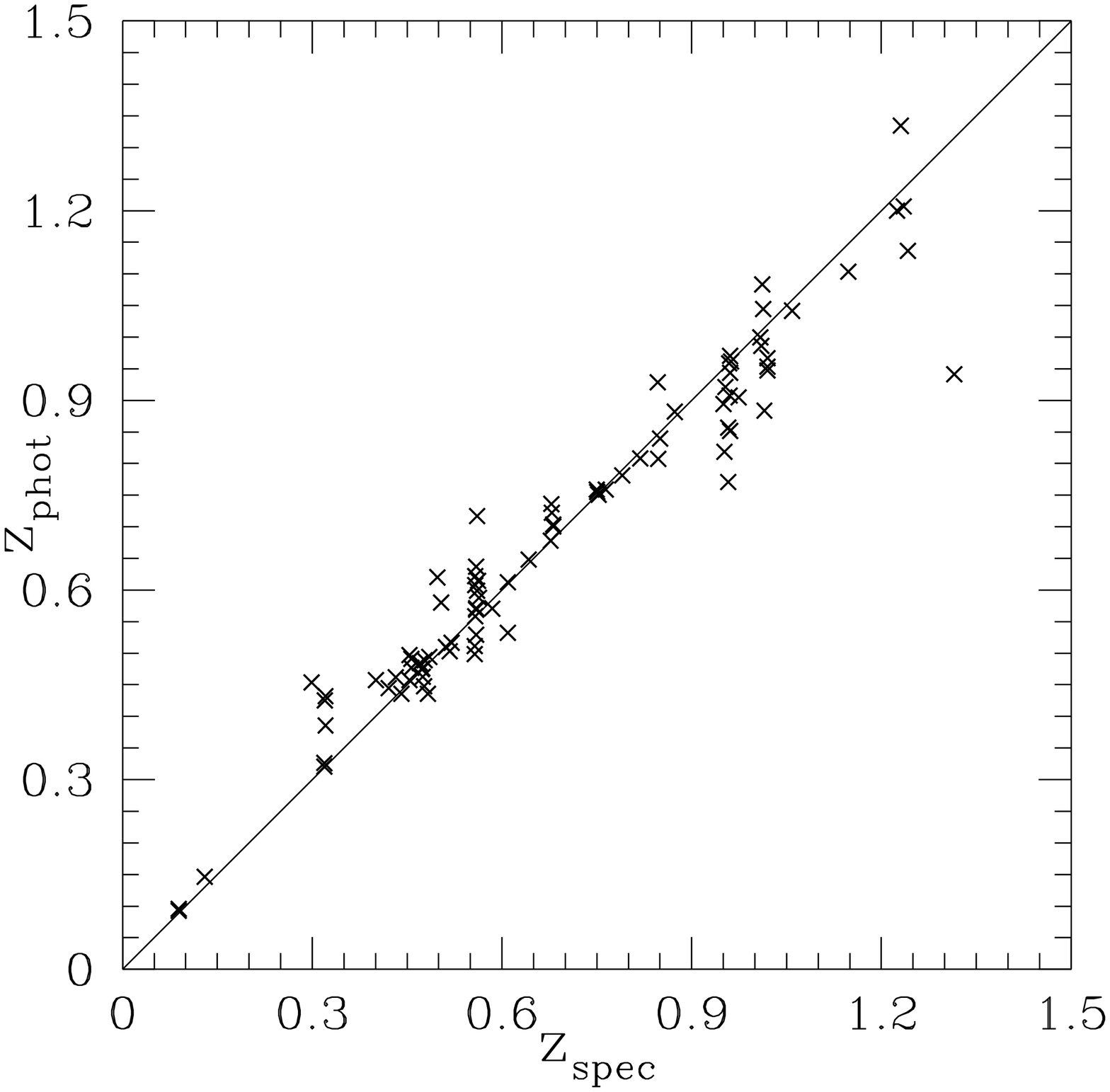}
\caption{Redshift limited scatter plot showing a close-up of Figure
\ref{zzfig}. \label{zzZfig}}
\end{figure}

\begin{figure}
\plotone{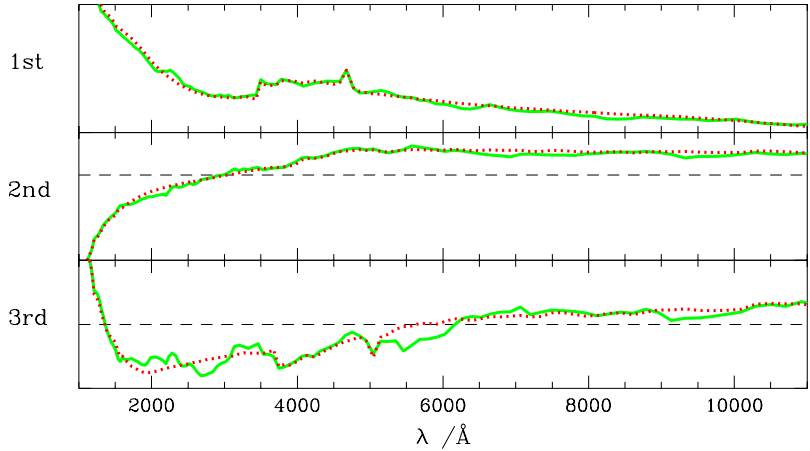}
\caption{These figures show the initial eigenspectra derived from the
Coleman, Wu \& Weedman spectra (dotted line) and the corrected KL spectra
after 30 iterations (solid line). The horizontal dashed line signals the zero
level. See Figure \ref{zzfig}. and \ref{zzZfig}. for related redshift scatter
plots. \label{eifig}}
\end{figure}

After  a few  iterations  the redshift  scatter   becomes tighter  while  the
eigentemplates just slightly change.  The comparison between the original and
resulting eigenspectra after  30 iterations (KL-30   ) can be  seen on Figure
\ref{eifig}.   This tiny difference between  them  was enough  to improve the
redshift scatter plot  by a  factor of  two  as  it  can  be  seen on  Figure
\ref{zzfig}.  and \ref{zzZfig}.   The  corresponding statistical errors  were
calculated  for two redshift ranges.  The  overall  statistics and a redshift
limited sample were evaluated.   Beyond the standard  root mean  square error
($\Delta_{\rm  rms}$)  the relative  deviation  of $(1+z)$ was  also computed
($\Delta_{\rm  rel}$) for comparison with  estimates found  by other authors.
Table \ref{scatter}.   contains the calculated error  estimates for  both the
CWW and KL-30 cases.

\begin{table}
\caption{Comparison of the errors in the fits based on the Coleman, Wu \&
Weedman and the KL (after 30 iterations) eigenspectra} \label{scatter}
\begin{center}
\begin{tabular}{cccc}
\multicolumn{1}{c}{Error\tablenotemark{}} &
Range & CWW & KL-30 \\
\tableline
\tableline
\multicolumn{1}{c}{$\Delta_{\rm rms}$\tablenotemark{}} & 
$z<6$ & 0.23 & 0.12 \\
\multicolumn{1}{c}{$\Delta_{\rm rel}$\tablenotemark{}} & 
$z<6$ & 0.079 & 0.042 \\
\tableline
\multicolumn{1}{c}{$\Delta_{\rm rms}$\tablenotemark{}} & 
$z<.8$ & 0.087 & 0.063 \\
\multicolumn{1}{c}{$\Delta_{\rm rel}$\tablenotemark{}} & 
$z<.8$ & 0.050 & 0.035 \\
\tableline
\end{tabular}
\end{center}
\end{table}

In order to test the  robustness of the  method we started the iteration from
arbitrary constant functions, as well.  After just a couple of iterations the
4000\AA{}  break was visible  and other important  features emerged soon from
scratch.

The limitation on  our current application  is the relatively small number of
galaxies with accurate multicolor photometry and redshifts in the Hubble Deep
Field. Consequently the third eigenspectrum appears to be  being over fit (we
run out of degrees of freedom). As our technique is constructed to be applied
to  any set  of  multicolor observations  we can incorporate  many multicolor
datasets from different sources (i.e.  we do not need  to train the  relation
for a given set of  filters or magnitude limits). With  the new generation of
multicolor  photometric and  spectroscopic   surveys nearing completion   we,
therefore,  expect the accuracy  of the derived spectral energy distributions
to improve dramatically in the near future.

\section{Conclusions}

We have shown  that  the current templates  used  in SED fitting  photometric
redshift estimation can be modified   to give better correlation between  the
actual photometry and  the template based simulated  fluxes.  We presented  a
robust  method that  creates  better eigenspectra  in  an  iterative way  and
converges very quickly.  The training  procedure  is very generic.  Different
catalogs  can  be  incorporated at     the same   time even  with   different
filtersets. It is also easy to extend the  method to involve measured spectra
if needed.

We showed that the  new modified eigentemplates  give us  a better  basis for
photometric  redshift estimation, even  though the  training procedure is not
fitting for the redshift, in fact the photometric redshift estimation is only
performed for monitoring purposes.

\acknowledgments 

IC and TB acknowledges partial support from the MTA-NSF grant no. 124 and the
Hungarian National Scientific Research Foundation  (OTKA) grant no.\ T030836,
AS  from   NASA LTSA (NAG53503), HST   Grant  (GO-07817-04-96A), MED from HST
(GO-07817-01-96A)  and     AJC    from   HST  (GO-07817-02-96A)   and    LTSA
(NRA-98-03-LTSA-039).


\begin{references}
\reference{} Baum, W.A., 1962, IAU Symposium No. 15, 390
\reference{} Budav\'{a}ri, T., Szalay, A.S., Connolly, A.J., Csabai, I. \&
Dickinson, M.E., 1999, in preparation, to be submitted to \aj
\reference{} Brunner, R.J., Connolly, A.J., Szalay, A.S., 1999, \apj, 
516:563-581 
\reference{} Bruzual, A.G. \& Charlot, S., 1993, \apj, 405, 538 
\reference{} Cohen, J.G., 1998, HDF Symposium (STScI symp. ser. 11), p. 52
\reference {} Coleman, G.D., Wu., C.-C. \& Weedman, D.W., 1980, \apjs, 43, 393
\reference{} Connolly, A.J., Csabai, I., Szalay, A.S., Koo, D.C., Kron, R.G. 
\& Munn, J.A., 1995a, AJ 110, 2655 
\reference{} Connolly A.J., Szalay A.S., Bershady M.A., Kinney A.L. \&
Calzetti D., 1995b, \aj, 110, 1071 
\reference{} Csabai, I., Connolly, A.J., Szalay, A.S. \& Budav\'{a}ri, T.,
1999, in press, submitted to \aj
\reference{} Dickinson, M., 1999, in ``After the Dark Ages:  When Galaxies
Were Young,'' eds. S. Holt and E. Smith, (AIP: Woodbury NY), p. 122
\reference{} Dickinson et al., 1999, in preparation
\reference{} Fern\'{a}ndez-Soto, A., Lanzetta, K.M. \& Yahil, A., 1999, \apj,
513:34-50
\reference{} Hogg et al., 1998, \apj, 499, 555
\reference{} Karhunen, H., 1947, Ann. Acad. Science Fenn, Ser. A.I. 37
\reference{} Lo\`{e}ve, M., 1948, Processus Stochastiques et Mouvement
Brownien, Hermann, Paris, France
\reference{} Williams, R.E. et al., 1996, \aj, 112, 1335
\end{references}
\end{document}